\documentclass{ws-ijmpa}

\begin{document}

\markboth{Paolo Benincasa}
{Sound waves in strongly coupled non-conformal gauge theory plasma}

\title{SOUND WAVES IN STRONGLY COUPLED NON-CONFORMAL GAUGE THEORY PLASMA}

\author{PAOLO BENINCASA}

\address{Department of Applied Mathematics, University of Western Ontario,\\
         Middlesex College, London, Ontario, N6A 5B7, Canada\\
         pbeninca@uwo.ca}

\maketitle

\vspace{1.25cm}

\begin{abstract}
Gauge/string correspondence provides an efficient method to investigate
gauge theories. In this talk we discuss the results of the paper (to
appear) by P. Benincasa, A. Buchel and A. O. Starinets, where the 
propagation of sound waves is studied in a strongly coupled non-conformal 
gauge theory plasma. In particular, a prediction for the speed of sound as 
well as for the bulk viscosity is made for the $\mathcal{N}=2^{*}$ gauge theory
in the high temperature limit. As expected, the results achieved show a 
deviation from the speed of sound and the bulk viscosity for a conformal 
theory. It is pointed out that such results depend on the particular gauge 
theory considered.

\keywords{Gauge/string correspondence; black holes in string theory; 
          correlation functions}
\end{abstract}

\ccode{PACS numbers: 11.25.Hf, 123.1K}

\section{Introduction}

This talk is based on the paper Ref.~\refcite{BBS} which analyzes the 
propagation of sound waves in a strongly coupled non-conformal gauge theory
plasma by using the gauge/string correspondence \cite{Mal}. This conjecture
establishes a duality between strongly coupled gauge theories and black hole 
supergravity models.

The hydrodynamics and the propagation of the sound waves were extensively 
studied for strongly coupled $\mathcal{N}=4$ superconformal Yang-Mills theory 
plasma\cite{PSS1}\cdash\cite{PSS3}. The known dispersion relation was 
reproduced for such a theory:
\begin{equation}
\omega(q)=v_{s} q -i\ \frac{2}{3}\
\frac{\eta}{s}\ \left(1+\frac{3\zeta}{4\eta}\right)\, \frac{q^2}{T}\,,
\end{equation}
where $v_{s}$ is the speed of sound, $\eta$ is the shear viscosity, $\zeta$ is
the bulk viscosity and $s$ is the entropy density.\footnote{It is to be 
noticed that the ratio between the shear viscosity and the entropy density is a
constant independent on the gauge theory considered\cite{BL1}\cdash\cite{Buc2}
:
\begin{equation}
\frac{\eta}{s}=\frac{1}{4\pi}.
\end{equation}
}
As a consequence of the conformal symmetry of the $\mathcal{N}=4$ gauge theory,
the speed of sound and the bulk viscosity turn out to assume the following
values:
\begin{equation}\label{vs}
v_{s}=\frac{1}{\sqrt{3}}, \qquad \zeta=0.
\end{equation}
It is easy to understand the above result for the speed of sound: for
conformal theories, the trace of the stress-energy tensor vanishes, so that 
the energy density and the pressure are related by $\mathcal{E}=3P$. The 
thermodynamics relation
\begin{equation}\label{thvs}
v_{s}^2=\frac{\partial P}{\partial\mathcal{E}}
\end{equation}
leads in a straightforward way to Eq. (\ref{vs}).

Non-conformal theories are instead expected to show different values for the
speed of sound $v_{s}$ and a non-vanishing bulk viscosity $\zeta$.
Our goal is the computation of these two quantities for non-conformal theories.
The gauge theory we study is the so-called $\mathcal{N}=2^{*}$ model: while it 
is difficult to make prediction for the real QCD because of the lack of a dual
string model, for the theory we analyze it is possible to understand some 
aspects of the dynamics from both the gauge theory side\cite{BPP,EJP} and the 
string theory one\cite{PW}.

The speed of sound is computed by using two different approaches in the high
temperature limit (what this means will be explained in the next section):
\begin{itemize}
\item the thermodynamics relation Eq. (\ref{thvs})
\item the hydrodynamic pole in the two-point correlation function for the
      stress-energy tensor. It is to be pointed out that the study of such 
      a pole also allows to compute the bulk viscosity $\zeta$
\end{itemize}
The speed of sound for the $\mathcal{N}=2^{*}$ gauge theory turns out to be 
less than the speed of sound for the conformal theories Eq. (\ref{vs}). 
As far as the bulk viscosity is concerned, it is non-vanishing, as expected. 
Moreover, both of these two quantities depend on parameters characteristic of 
the gauge theory considered: this means that speed of sound and bulk viscosity 
are gauge-theory specific.

\section{The non-conformal gauge theory: the $\mathcal{N}=2^{*}$ model}

We start by a quick review of the $\mathcal{N}=2^{*}$ gauge theory. As was 
already mentioned, we consider such a model because it can be analyzed both 
from the gauge theory perspective and from the dual supergravity one. This 
allows to compare the results and they are generally found in agreement with 
each other.

The $\mathcal{N}=2^{*}$ theory is the mass deformed $\mathcal{N}=4$ 
supersymmetric Yang-Mills theory. In the $\mathcal{N}=1$ language, the 
$\mathcal{N}=4$ supersymmetric Yang-Mills theory contains a vector multiplet 
$V$ and three chiral multiplets $\Phi, Q, \tilde{Q}$. Moreover, it is
endowed with the superpotential
\begin{equation}\label{suppot}
W=\frac{2\sqrt{2}}{g_{YM}^{2}}\:tr\left(\left[Q,\tilde{Q}\right]\Phi\right).
\end{equation}

If a mass deformation is introduced in the theory so that the bosonic and
fermionic components of $Q, \tilde{Q}$ receive the same mass $m$, the 
superpotential assumes the form
\begin{equation}\label{suppot2}
W=\frac{2\sqrt{2}}{g_{YM}^{2}}tr\left(\left[Q,\tilde{Q}\right]
              \Phi\right)+\frac{m}{g_{YM}^{2}}\left(tr\{Q^2\} + 
              tr\{\tilde{Q^2}\}\right)
\end{equation}
and the $\mathcal{N}=4$ supersymmetry is softly broken to $\mathcal{N}=2$ 
with $V, \Phi$ as $\mathcal{N}=2$ vector multiplet and $Q, \tilde{Q}$ 
as a hypermultiplet.

If the bosonic and fermionic components of $Q, \tilde{Q}$ receive 
masses $m_{b}$ and $m_{f}$ respectively with $m_{b}\neq m_{f}$, the 
supersymmetry is completely broken. 

An explicit dual supergravity realization exists at large 't Hooft coupling
$g_{YM}^2 N \gg 1$ \cite{PW}. In addition, the black brane geometry is 
analytically known in the high temperature limit\cite{BL2}.

The supergravity realization of the finite temperature $\mathcal{N}=2^{*}$ 
gauge theory is described by the following five-dimensional effective
action:
\begin{equation}\label{effact}
 S=\frac{1}{4\pi G_{5}}\int_{{\mathcal{M}}_{5}}d^{5}\xi
             \sqrt{-g}\left[\frac{1}{4}R-3(\partial\alpha)^2-(\partial\chi)^2-
             \mathcal{P}\right],
\end{equation}
where:
\begin{equation}\label{P}
\mathcal{P}=\frac{1}{16}\left[\frac{1}{3}
             \left(\frac{\partial W}{\partial\alpha}\right)^{2}+
             \left(\frac{\partial W}{\partial\chi}\right)^{2}\right]-
             \frac{1}{3}W^{2},
\end{equation}
and 
\begin{equation}\label{W}
W=-e^{-2\alpha}-\frac{1}{2}e^{4\alpha}\cosh{2\chi} .
\end{equation}
It is to be pointed out that the coefficients of the leading asymptotic \
behaviour near the boundary for the two supergravity scalar fields $\alpha$ 
and $\chi$ are respectively related to the bosonic and fermionic mass 
parameters of the gauge theory. 

From the effective action (\ref{effact}), it is quite straightforward to obtain
the equations of motion 
with the metric {\it ansatz}:
\begin{equation}\label{metric}
ds^{2}=-c_{1}^{2}(r)dt^{2}+c_{2}^{2}(r)d\vec{x}^{2}+c_{3}^{2}(r)dr^{2}
\end{equation}
\begin{eqnarray}\label{eom}
0 &=& \alpha''+\alpha'\ \left[\ln\frac{c_1c_2^3}{c_3}\right]'-
      \frac 16 c_3^2\frac{\partial\mathcal{P}}{\partial\alpha} \nonumber \\
0 &=& \chi''+\chi'\ \left[\ln\frac{c_1c_2^3}{c_3}\right]'-
      \frac 12 c_3^2\frac{\partial\mathcal{P}}{\partial\chi} \nonumber \\
0 &=& c_1''+c_1'\ \left[\ln\frac{c_2^3}{c_3}\right]'+\frac{4}{3}c_1 c_3^2 
      \mathcal{P}   \nonumber \\
0 &=& c_2''+c_2'\ \left[\ln\frac{c_1c_2^2}{c_3}\right]'+\frac{4}{3}c_2 c_3^2 
      \mathcal{P} .
\end{eqnarray}
There is also a first order constraint
\begin{equation}\label{constr}
0=\left(\alpha'\right)^2+\frac{1}{3} \left(\chi'\right)^2-\frac{1}{3} c_3^2 
  \mathcal{P}-\frac{1}{2} [\ln c_2]'[\ln c_1 c_2]' .
\end{equation}

The background equations (\ref{eom}), together with the constraint 
(\ref{constr}), can be analytically solved\cite{BL2} in the limit 
of high temperature, i.e. when the temperature $T$\footnote{The temperature
in the gauge theory corresponds in the supergravity side to the Hawking 
temperature of the black hole background.} is much larger than the
mass parameters $\{m_{b}, m_{f}\}$:
\begin{equation}\label{HT}
\left(\frac{m_{b}}{T}\right)^2\ll 1, \qquad \frac{m_{f}}{T}\ll 1.
\end{equation}
This fact will be crucial for our computation.

\section{The speed of sound from thermodynamics}

The idea for the computation of the speed of sound from thermodynamics is
quite simple:
\begin{itemize}
\item compute the one-point correlation function for the stress-energy
      tensor $\Big< T_{\mu\nu}\Big>$;
\item extract the pressure $P$ and the energy density $\mathcal{E}$ from
      $\Big< T_{\mu\nu}\Big>$;
\item apply the thermodynamics relation (\ref{vs})
\end{itemize}
The one-point correlation function for the stress-energy tensor, which is 
defined at the boundary of the spacetime ${\mathcal{M}}_{5}$, turns out to
be infinite, so it requires to be regularized and renormalized\cite{BK}\cdash
\cite{Buc}: the spacetime ${\mathcal{M}}_{5}$ can be 
considered as foliated along the radial coordinate $r$ in timelike surfaces 
$\partial{\mathcal{M}}_{r}$ (they are homomorphic to the boundary 
$\partial{\mathcal{M}}_{5}$ of ${\mathcal{M}}_{5}$) whose metric is evaluated 
at the boundary. 
The renormalized stress-energy tensor is obtained by evaluating at the boundary
$\partial{\mathcal{M}}_{5}$ the variation, with respect the metric of 
$\partial{\mathcal{M}}_{r}$, of an action which is the sum of (\ref{effact}),
the standard Gibbons-Hawking term and the counterterm action $S_{c}$ 
\footnote{The expression of the counterterm action $S_{c}$ can be found in 
Refs.~\refcite{Buc,BBS}.} (the last
contribution is introduced to achieve a finite result):
\begin{equation}\label{1pt}
T_{\mu\nu}=T_{\mu\nu}^{\small{reg}}\Big |_{\partial{\mathcal{M}}_{5}},
\end{equation}
where:
\begin{equation}\label{1ptreg}
 T_{\mu\nu}^{\small{reg}}=\frac{1}{8\pi G_{5}}
         \left[-\theta_{\mu\nu}+\theta\gamma_{\mu\nu}\right]+
         \frac{2}{\sqrt{-\gamma}}
         \frac{\delta S_{\small{c}}}{\delta\gamma^{\mu\nu}}.
\end{equation}
Here, $\gamma_{\mu\nu}$ is the metric of $\partial{\mathcal{M}}_{r}$, 
$\theta_{\mu\nu}$ is the extrinsic curvature, $\theta$ is the trace of
$\theta_{\mu\nu}$ ($\theta=\theta_{\mu\nu}\gamma^{\mu\nu}$).

Thus, the energy density (ADM mass) can be extracted from (\ref{1pt}):
\begin{equation}\label{ADM}
\mathcal{E}=\sqrt{\sigma}N_{\Sigma}u^{\mu}u^{\nu}T_{\mu\nu},
\end{equation}
where $u^{\mu}$ is the unit vector normal to a spacelike surface $\Sigma$ in
$\partial{\mathcal{M}}_{5}$, $\sigma$ is the determinant of the induced metric
on $\Sigma$, $N_{\Sigma}$ is the norm of the timelike Killing vector in 
(\ref{metric}).
Moreover, since $\Sigma$ is homogeneous and isotropic, the pressure is given
by:
\begin{equation}\label{press}
P=\sqrt{\sigma}N_{\Sigma}T_{x_{1}x_{1}}
              \gamma^{x_{1}x_{1}}.
\end{equation}

From the asymptotic expansions of the solutions of the equations (\ref{eom}) 
and (\ref{constr}) (for more computational details, see Ref.~\refcite{BBS}) 
and from (\ref{ADM}) and 
(\ref{press}), we obtain in the high temperature limit at the lowest order in
$\left(\frac{m_{b}}{T}\right)^2$ and $\left(\frac{m_{f}}{T}\right)$:
\begin{equation}\label{Ef}
\mathcal{E}=
              \frac{3}{8}\pi^{2}N^{2}T^{4}
              \left[\mathbf{1}+\frac{1}{9\pi^{4}}
              \left(\ln{T\pi}-1\right)
              \mathbf{\left(\frac{m_{b}}{T}\right)^{4}}-
              \frac{2[\Gamma(\frac{3}{4})]^{4}}{3\pi^{4}}
              \mathbf{\left(\frac{m_{f}}{T}\right)^{2}}
              +\ldots\right]
\end{equation}
\begin{equation}\label{Pf}
P=
              \frac{1}{8}\pi^{2}N^{2}T^{4}
              \left[\mathbf{1}-\frac{1}{3\pi^{4}}\ln{T\pi}\:
              \mathbf{\left(\frac{m_{b}}{T}\right)^{4}}-
              \frac{2[\Gamma(\frac{3}{4})]^{4}}{\pi^{4}}
              \mathbf{\left(\frac{m_{f}}{T}\right)^{2}}
              +\ldots\right].
\end{equation}
It is also possible to compute the entropy density\cite{BL2}:
\begin{equation}\label{s}
s=\frac{1}{2}\pi^2 N^2 T^{3}
  \left[1-\frac{1}{12\pi^4}\left(\frac{m_{b}}{T}\right)^{4}-
          \frac{[\Gamma(\frac{3}{4})]^{4}}{\pi^4}
              \left(\frac{m_{f}}{T}\right)^{2}+\ldots\right]
\end{equation}
and verify that (\ref{Ef}), (\ref{Pf}) and (\ref{s}) satisfy the relation
\begin{equation}\label{rel}
\mathcal{E}-Ts=-P
\end{equation}
as expected.

It is now easy to apply the formula (\ref{vs}) and obtain the leading 
correction to the speed of sound:
\begin{equation}\label{vsf}
v_{s}=\frac{1}{\sqrt{3}}\left[\mathbf{1}-
                  \frac{1}{18\pi^{4}}
                  \mathbf{\left(\frac{m_{b}}{T}\right)^{4}}-
                  \frac{[\Gamma(\frac{3}{4})]^{4}}{3\pi^{4}}
                  \mathbf{\left(\frac{m_{f}}{T}\right)^{2}}
                  +\ldots\right].
\end{equation}
Some comments about (\ref{vsf}), (\ref{Ef}) and (\ref{Pf}) are mandatory. It
is to be pointed out that in absence of mass deformations, the correct
results for conformal gauge theories are reproduced (they are given by the
zeroth-order terms): this was expected since, without the mass deformation,
the $\mathcal{N}=2^{*}$ theory reduces to the $\mathcal{N}=4$ superconformal
Yang-Mills. Moreover, all these expressions show corrections which are 
mass-parameter dependent: this implies that such results are not universal, but
they depend on the particular gauge theory. As a final remark, the lowest order
corrections to the speed of sound are negative in sign: the speed of sound
for the $\mathcal{N}=2^{*}$ theory is predicted to be less than the speed of
sound for a conformal field theory.

\section{The speed of sound from the two-point correlation function of the
         stress-energy tensor}

As is known\cite{AGMOO}, the correlation functions of a gauge theory can be 
reproduced from supergravity by studying the fluctuations of the background 
geometry:
\begin{eqnarray}\label{fluct}
g_{\mu\nu} &\longrightarrow& g_{\mu\nu}^{(0)}+h_{\mu\nu}\\
\alpha &\longrightarrow& \alpha^{(0)}+\alpha^{(1)}\\
\chi &\longrightarrow& \chi^{(0)}+\chi^{(1)}
\end{eqnarray}
where $\{g_{\mu\nu}^{(0)}, \alpha^{(0)}, \chi^{(0)}\}$ satisfy the background
equations of motions and $\{ h_{\mu\nu}, \alpha^{(1)}, \chi^{(1)}\}$ are the
fluctuations of the supergravity fields. We choose the following gauge:
\begin{equation}\label{gauge}
h_{r\mu}=0
\end{equation}
and assume that the perturbations depend only on $(t, x_{3}, r)$: there is an
$O(2)$ rotational invariance in the $xy$-plane which allows them to be 
classified by the spin:
\begin{itemize}
\item spin 2: $\{h_{x_{1} x_{2}}\}$; $\{h_{x_{1} x_{1}}-h_{x_{2} x_{2}}\}$
\item spin 1: $\{h_{tx_{1}}, h_{x_{1} x_{3}}\}$; 
              $\{h_{t x_{2}}, h_{x_{2} x_{3}}\}$
\item spin 0: $\{h_{tt}, h_{x_{1} x_{1}}+h_{x_{2} x_{2}}, h_{x_{3}x_{3}}, 
                 h_{tx_{3}}, \alpha^{(1)}, \chi^{(1)}\}$.
\end{itemize}
These sets of fluctuations decouple from each other. The spin 2 
perturbations separately describe the scalar modes; the two sets of spin 1 
fluctuations separately allow to compute the 
correlation functions showing the diffusion pole; the spin 0 set contains the
sound modes\footnote{More about this classification is explained in 
Ref.~\refcite{PSS2}.}. 
We are, therefore, interested in the scalar perturbations.

The correlation functions are computed by using the Minkowski prescription 
proposed in Ref.~\refcite{SS}: in order to obtain retarded Green functions, we 
require, as one of the boundary conditions,  that the fluctuations correspond 
to incoming wave at the horizon. Such a boundary condition is to be imposed 
only on the physical modes: as shown in Ref.~\refcite{PSS3}, pure gauge 
solutions are also present, but they can be eliminated\cite{Sal,BBS}.

It is now possible to analyze the physical fluctuations in the hydrodynamic 
limit, i.e. in the low frequency and momentum limit with their ratio kept 
constant:
\begin{equation}\label{hydro}
\omega\longrightarrow 0,\qquad q\longrightarrow 0, 
                        \qquad \frac{\omega}{q}=const.
\end{equation}
In this way we obtain some analytical equations which require numerical 
techniques to be solved\footnote{All the computational steps are described in 
Ref.~\refcite{BBS}.}.

For convenience, the dispersion relation for the hydrodynamic pole is 
parametrized in the following way:
\begin{eqnarray}\label{hydropole}
\omega &=& \frac{q}{\sqrt{3}}
              \left[1+\beta_{1}^{v}\frac{1}{576\pi^2}
               \left(\frac{m_{b}}{T}\right)^{4}+
               \beta_{2}^{v}\frac{[\Gamma(\frac{3}{4})]^{4}}{4\pi^3}
               \left(\frac{m_{f}}{T}\right)^{2}
              \right] \nonumber \\
       &-&    i\frac{q^{2}}{3}
              \left[1+\beta^{\Gamma}_{1}\frac{1}{576\pi^2}
               \left(\frac{m_{b}}{T}\right)^{4}+
               \beta_{2}^{\Gamma}\frac{[\Gamma(\frac{3}{4})]^{4}}{4\pi^3}
               \left(\frac{m_{f}}{T}\right)^{2}
              \right].
\end{eqnarray}
It is evident how the coefficients $\{\beta_{1}^{v}, \beta_{2}^{v}\}$ are
related to the speed of sound, while $\{\beta_{1}^{\Gamma}, 
\beta_{2}^{\Gamma}\}$ provide the dispersion. The numerical techniques 
discussed in Ref.~\refcite{BBS} lead to the following results:
\begin{itemize}
\item \begin{equation}\label{coeffv}
          \beta_{1}^{v}=-\frac{32}{\pi^2}\cdot 1.00000(1),
          \qquad
          \beta_{2}^{v}=-\frac{4}{3\pi}\cdot 0.9999(5)
      \end{equation}
\item \begin{equation}\label{coeffd}
          \beta_{1}^{\Gamma}= 8.001(8),
          \qquad
          \beta_{2}^{\Gamma}= 0.9672(1),
       \end{equation}
\end{itemize}
where the number between brackets indicates an error in the corresponding 
digit.
\footnote{These results are obtained by a general purpose software. The 
          precision can be improved only by using a dedicated program.}

A comparison between the coefficients (\ref{coeffv}) and the value for the 
speed of sound obtained by the thermodynamics approach shows that the results
are in perfect agreement. This is a non-trivial check for our prediction.

As far as the coefficients (\ref{coeffd}) are concerned, they allow to make
a prediction for the bulk viscosity:
\begin{equation}\label{bulk}
\zeta=\frac{\pi N^{2}T^{3}}{6}
              \left[\beta_{1}^{\Gamma}\frac{1}{576\pi^2}
              \left(\frac{m_{b}}{T}\right)^{4}+
              \beta_{2}^{\Gamma}\frac{[\Gamma(\frac{3}{4})]^{4}}{4\pi^3}
              \left(\frac{m_{f}}{T}\right)^{2}+\ldots\right].
\end{equation}

As expected, the $\mathcal{N}=2^{*}$ gauge theory is characterized by a 
non-vanishing bulk viscosity. As in the case of the speed of sound, it is 
possible to argue from the expression (\ref{bulk}) that, in absence of mass 
deformation, the known result for the conformal theories is obtained. Moreover,
the corrections introduced by the mass deformation are mass-parameter 
dependent: the bulk viscosity is gauge-theory specific as well.

It is possible to rewrite (\ref{bulk}) in the following way:
\begin{equation}\label{bsv}
\frac{\zeta}{\eta} = \left[\beta_{1}^{\Gamma}\frac{1}{432\pi^2}
              \left(\frac{m_{b}}{T}\right)^{4}+
              \beta_{2}^{\Gamma}\frac{[\Gamma(\frac{3}{4})]^{4}}{3\pi^3}
              \left(\frac{m_{f}}{T}\right)^{2}+\ldots\right].
\end{equation}
From the expressions of the speed of sound (\ref{vsf}) and of the bulk
viscosity (\ref{bsv}), it is possible to argue that, at least in the 
high temperature limit, the ratio of the bulk viscosity to shear viscosity
is proportional to the deviation of the speed of sound from its value in the
conformal theories:
\begin{equation}\label{ratio}
\frac{\zeta}{\eta}\simeq - \kappa \, \left(v_{s}^{2}-\frac{1}{3}\right),
\end{equation}
where:
\begin{equation}\label{kappa}
\kappa \approx 
 \left\{
  \begin{array}{l}
   4.935 \qquad \mbox{for } m_{f}=0 \\
   4.558 \qquad \mbox{for } m_{b}=0.
  \end{array}
 \right.
\end{equation}
Such a result disagrees with the relation predicted in Ref.~\refcite{HST,HS} 
and already criticized in Ref.~\refcite{JY}:
\begin{equation}\label{wrong}
\zeta \sim \eta\left(v_{s}^{2}-\frac{1}{3}\right)^{2}.
\end{equation}

\section{Conclusion}

The gauge/string correspondence allowed us to study the sound waves for a
non-conformal gauge theory (the $\mathcal{N}=2^{*}$ model).
In particular, the speed of sound for such a theory was obtained by two 
different computations: from thermodynamics by (\ref{thvs}) relating speed of 
sound, pressure and density of energy, and from the hydrodynamic pole. The
results are found to be in agreement with each other. Moreover, the speed
of sound for the $\mathcal{N}=2^{*}$ gauge theory turned out to be less than
the speed of sound for the conformal theories: in the high temperature regime 
the mass deformations introduced negative corrections.

Moreover, a prediction for the bulk viscosity for
a non-conformal gauge theory was made: as expected, the value found was 
non-vanishing.

Both the speed of sound and the bulk viscosity turned out to be non-universal
quantities: the corrections depend on the mass-parameter of the particular
theory, so that different values of such quantities correspond to different
non-conformal strongly coupled gauge theories.

We hope that this dual supergravity analysis of plasma transport coefficients
can be useful both in a future analysis of such coefficients for QCD and in 
hydrodynamic models at RHIC.

\section*{Acknowledgements}

I would like to thank Amir Fariborz for the stimulating environment created at
the MRST-2005 conference. It is a pleasure to thank Alex Buchel, who has 
prompted my interest in the subject, for many explanations and for reviewing 
the manuscript. I am also grateful to Andrei Starinets for further comments on
the manuscript and for many valuable discussions and to Francesco Sannino for 
encouragement.

%
%

\end{document}